# Gaussian Process-Based Scalar Field Estimation in GPS-Denied Environments

Muzaffar Qureshi[1]  Tochukwu Elijah Ogri[1]  Humberto Ramos[1]  Zachary I. Bell[2]  Rushikesh Kamalapurkar[1]

*Abstract*— This paper presents a methodology for an autonomous agent to map an unknown scalar field in GPS-denied regions. To reduce localization errors, the agent alternates between GPS-enabled and GPS-denied areas while collecting measurements. User-defined error bounds determine the dwell time in each region. A switching trajectory is then designed to ensure field measurements in GPS-denied regions remain within the specified error limits. A Lyapunov-based stability analysis guarantees bounded error trajectories while tracking the desired path. The methodologys effectiveness is demonstrated through simulations, with an error analysis comparing the GP-predicted scalar field model to the actual field.

## I. INTRODUCTION

Mapping unknown scalar fields is critical in numerous scientific and engineering applications, including environmental monitoring (e.g., mapping level of pollution), search and rescue operations (e.g., mapping radiation levels), and autonomous underwater exploration (e.g., mapping thermal gradients) [1]–[3]. In many real-world applications, however, autonomous agents tasked with mapping scalar fields must operate in GPS-denied environments where the agent's state feedback is unavailable, like in underwater exploration and stealthy mapping of hostile radar fields to avoid GPS detection [4]–[6]. In these environments, agents often rely on inertial and relative measurements to estimate their trajectories, resulting in inevitable errors over time.

Robotic exploration in GPS-denied environments using alternative sensors, in addition to relative and/or inertial measurements, has received extensive attention in the literature [7]–[9]. Often, visual or LiDAR-based simultaneous localization and mapping (SLAM) techniques are used. These approaches leverage feature recognition for localization and loop-closures (recognition of previously visited places) to minimize localization's drift [10]–[13]. However, these methods depend on distinct and recognizable environmental features to provide accurate state estimates. In feature-sparse environments such as underwater, desert, or foggy conditions their performance degrades significantly, making reliable state estimation difficult.

Given these limitations, an alternative approach is to mitigate localization errors by intermittently accessing GPS-enabled regions within the domain. This switched systems framework approach strategically alternates between GPS-denied and GPS-enabled regions [14], [15]. This approach allows the agent to perform tasks in GPS-denied regions while periodically correcting localization drift, thereby improving the overall agent's state estimation accuracy and mission reliability.

The idea of this switched system framework is that the agent operates in GPS-denied regions but intermittently visits GPS-enabled regions to keep the state estimation error bounded. Chen et al. [15] derived dwell time conditions through Lyapunov-based stability analysis, which govern the minimum and maximum time the agent can spend in each region based on the user-defined acceptable error bounds. This method is particularly effective in structured environments where GPS access is intermittent, such as urban canyons, subterranean tunnels with occasional surface access, or large indoor facilities with limited satellite visibility. By intelligently balancing navigation autonomy (GPS-denied regions) and localization correction (GPS-enabled regions), the switched systems framework provides a robust and adaptable solution for state estimation in challenging environments.

This paper addresses the challenge of scalar field mapping in GPS-denied environments, where the absence of state feedback complicates the agent's localization and trajectory tracking. We propose a switched systems-based framework approach that leverages intermittent access to GPS-enabled regions to bound localization errors while enabling efficient field mapping when GPS is unavailable. We build on the dwell time conditions of Chen et al. [15], to derive and extend the switched systems framework for scalar field mapping and ensure optimal time allocation between GPS-denied and GPS-enabled regions. This enables the design of a feasible switching trajectory that visits the desired measurement locations in GPS-denied regions while keeping state estimation errors bounded. By integrating switched systems theory with autonomous field mapping, the result is an adaptive solution for robust exploration in GPS-denied settings. , with applications in environmental monitoring, planetary exploration, and autonomous navigation in complex terrains.

## II. PROBLEM FORMULATION

In this paper, we assume that a sensor-equipped agent operating in a domain $\mathcal{X} \subset \mathbb{R}^p$ is tasked with visiting a set of measurement locations and creating a map of a scalar field denoted $h : \mathcal{X} \to \mathbb{R}$. The domain $\mathcal{X}$ is divided into two distinct regions, $\mathcal{X}_a$, where the agent can access its

This research is supported by the Air Force Research Laboratories under contract numbers FA8651-23-1-0006 and FA 8651-24-1-0019. Any opinions, findings, or recommendations in this article are those of the author(s) and do not necessarily reflect the views of the sponsoring agency.
[1] Department of Mechanical and Aerospace Engineering, University of Florida, email: {muzaffar.qureshi, tochukwu.ogri, jramoszuniga, rkamalapurkar} @ufl.edu.
[2] Air Force Research Laboratories, Florida, USA, email: zachary.bell.10@us.af.mil.

full state for feedback, and $\mathcal{X}_u$, where state feedback is not available. The two regions are assumed to intersect only at the boundary, that is, $\mathcal{X}_a \cup \mathcal{X}_u = \mathcal{X}$ and $\mathcal{X}_a \cap \mathcal{X}_u = \partial \mathcal{X}_a \cap \partial \mathcal{X}_u \neq \emptyset$, where $\partial \mathcal{X}$ represent the boundary of the set $\mathcal{X}$.

The measurement locations are assumed to lie in $\mathcal{X}_u$, and are denoted by $\mathbf{X} := \{\mathbf{x}_i\}_{i=1}^m$, where $\mathbf{x}_i \in \mathcal{X}_u, \forall i = 1, 2, \ldots m$, where $m \in \mathbb{N}$. The corresponding field measurements are represented as a vector $\mathbf{Z} := \{z_i\}_{i=1}^m$, where $z_i = h(\mathbf{x}_i) + \epsilon$, where $\epsilon$ is the zero-mean Gaussian noise, i.e., $\epsilon \sim \mathcal{N}(0, \sigma^2)$, with variance $\sigma^2$.

The following control-affine system describes the dynamics of the agent

$$\dot{x}(t) = f(t, x) + u(t) + g(t), \tag{1}$$

where $x \in \mathbb{R}^n$ represents the state of the agent, $f : [0, \infty) \times \mathbb{R}^n \to \mathbb{R}^n$ denotes the known drift dynamics, $u : \mathbb{R} \to \mathbb{R}^n$ is the control input, and $g : [0, \infty) \to \mathbb{R}^n$ represents the time-varying unmodeled dynamics causing the actual states to deviate from desired states. The following assumption is required to guarantee the existence and uniqueness of solutions (1).

*Assumption 1:* The function $f(t, x)$ is piecewise continuous with respect to $t$ and locally Lipschitz continuous with respect to $x$. Furthermore, $f(t, 0) = 0$ for all $t \geq 0$. Additionally, the unmodeled dynamics $g(t)$ satisfies $0 < \|g(t)\| \leq \bar{g}$ for all $t \geq 0$, where $\bar{g} \in \mathbb{R}_{\geq 0}$ is a known upper bound.

Since state feedback is available in $\mathcal{X}_a$, a feedback controller can be designed using existing control techniques like adaptive control. Therefore, this paper focuses on developing a trajectory that enables the agent to reach measurement locations in $\mathbf{X}$ without state feedback. Due to the unmodeled dynamics, state estimation errors accumulate when the agent operates in $\mathcal{X}_u$ for an extended period. Therefore, the main objective of this paper is to design a trajectory that enables the agent to visit the measurement locations $\mathbf{X}$ while keeping the state estimation error within desired bounds.

## III. OVERVIEW

The proposed methodology is to integrate a switching trajectory framework with the scalar field mapping task to control the error between the estimated states and actual states whenever the agent is in the GPS-denied region $\mathcal{X}_u$. The actual states of the agent are affected by unmodeled dynamics $g(t)$, and due to loss of state feedback in $\mathcal{X}_u$, the state estimation error tends to grow. To keep these state estimation errors bounded, the agent must visit $\mathcal{X}_a$ at regular time intervals to attain state feedback and correct its state estimates. To design the desired switching trajectory $\bar{x}_d$, an observer is needed to estimate the current states of the agent in the GPS-denied region $\mathcal{X}_u$. The following error definitions are required to facilitate the observer design.

Let state estimation error $e_1 : [0, \infty) \to \mathbb{R}^n$ between the estimated state $\hat{x}(t) \in \mathbb{R}^n$ and the actual state $x(t)$ of the system in (1) be defined as

$$e_1(t) := x(t) - \hat{x}(t). \tag{2}$$

Let the trajectory tracking error $e_2 : [0, \infty) \to \mathbb{R}^n$, which is the difference between estimated state $\hat{x}$ of the agent and the desired state $\bar{x}_d(t) \in \mathbb{R}^n$, be defined as

$$e_2(t) := \hat{x}(t) - \bar{x}_d(t). \tag{3}$$

The total error is defined as $e(t) := [e_1(t)^\top, e_2(t)^\top]^\top \in \mathbb{R}^{2n}$. Note that $e_1$ is only measurable when the agent is operating in $\mathcal{X}_a$, while $e_2$ is measurable at all times. Let $V : \mathbb{R}^{2n} \to \mathbb{R}$ be a candidate Lyapunov function defined as

$$V(e(t)) = \frac{1}{2} e_1^T e_1 + \frac{1}{2} e_2^T e_2, \quad \forall t \in [0, \infty). \tag{4}$$

The desired trajectory $\bar{x}_d$ depends on two user-selected bounds $V_u > 0$ and $V_l > 0$. The upper bound $V_u$ is relevant to the segment of the trajectory in $\mathcal{X}_u$, where the estimation errors grow, and the Lyapunov function tends to increase. By bounding the Lyapunov function from above, we also bound the maximum allowable time the agent can stay in $\mathcal{X}_u$. On the other hand, the lower bound $V_l$ is relevant to $\mathcal{X}_a$ which defines the minimum time the agent must stay in $\mathcal{X}_a$ before exiting into $\mathcal{X}_u$. The idea is to use a Lyapunov-based argument to bound the rate of increase of $V$ from above when the agent is in $\mathcal{X}_u$. Similarly, we bound the decrease of $V$ from below. These bounds are then used to find an upper bound on the time the agent can spend in $\mathcal{X}_u$ and a lower bound on the time the agent must spend in $\mathcal{X}_a$ to ensure $V_l \leq V(e(t)) \leq V_u$ for all $t$. These time bounds, along with the error bounds, are then used to design $\bar{x}_d$ such that, despite tracking errors, the agent's trajectory satisfies the time bounds.

In the following development, a Gaussian Process (GP) is employed to learn the scalar field based on the measurements during the experiment.

## IV. GAUSSIAN PROCESS REGRESSION

This section presents a Gaussian Process (GP) regression framework for mapping the scalar field $h$ based on the measurements $\mathbf{Z}$ collected by the mobile agent at measurement locations $\mathbf{X} = \{\mathbf{x_1}, \mathbf{x_2}, \ldots \mathbf{x_m}\}$. GP regression, a probabilistic model for learning non-linear functions [16], generates the scalar field map as a random function $\bar{h}$ characterized as

$$\bar{h} \sim \text{GP}(\mu, \mathbf{k}), \tag{5}$$

where $\mu : \mathbb{R}^p \to \mathbb{R}$ is the mean function and $k : \mathbb{R}^p \times \mathbb{R}^p \to \mathbb{R}$ is the covariance function. Among the common choice of kernel functions, we select a squared exponential covariance kernel defined as

$$k(\mathbf{x}, \mathbf{x}') := \alpha^2 \exp\left(-\frac{\|\mathbf{x} - \mathbf{x}'\|^2}{2\beta^2}\right), \tag{6}$$

where $\alpha$ represents the amplitude (signal variance), and $\beta$ is the length scale controlling the smoothness of the function. $\mathbf{x}$ and $\mathbf{x}'$ represent any two points in the domain. The squared exponential kernel is chosen due to its smoothness properties and ability to effectively model continuous and differentiable functions.

Given any query location $\mathbf{x}_e$, the training kernel matrix $K_{m-m} \in \mathbb{R}^{m \times m}$, the test-train kernel matrix $K_{e-m} \in$

$\mathbb{R}^{1\times m}$, and the testing kernel matrix $K_{e-e} \in \mathbb{R}$ can now defined based on the elements of training and testing locations set and the chosen kernel function. Mathematically, these matrices can be written as

$$K_{m-m}(j,k) := k(\mathbf{x}_j, \mathbf{x}_k), \quad (7)$$

$$K_{e-e} := k(\mathbf{x}_e, \mathbf{x}_e), \quad (8)$$

$$K_{e-m}(k) := k(\mathbf{x}_e, \mathbf{x}_k), \quad (9)$$

Given a set of measurement locations $\mathbf{X}$ and corresponding measurements $\mathbf{Z}$ at time $\tau$, the GP model is used to predict the mean and covariance evaluated at any given test point $\mathbf{x}_e$. The mean $\bar{\mathbf{x}}_e$ representing the predicted values of the GP model at $\mathbf{x}_e$, is given by

$$\bar{\mathbf{x}}_e(\tau) = K_{e-m}(\tau) \cdot (K_{m-m}(\tau))^{-1} \mathbf{Z}(\tau). \quad (10)$$

The covariance matrix $\mathbf{S}$, which captures the uncertainty in the GP prediction, is computed as Squeezed equation

$$\mathbf{S}(\tau) = K_{e-e} - K_{e-m}(\tau) \cdot (K_{m-m}(\tau))^{-1} \cdot (K_{e-m}(\tau))^\top. \quad (11)$$

The subsequent sections focus on the design of a trajectory that enables the agent to visit all desired measurement locations in $\mathbf{X}$ and collect the corresponding field values $\mathbf{Z}$, ensuring efficient exploration while maintaining accurate field estimation.

## V. Controller Design and Observer Update Laws

This section presents the design of a state observer and a stabilizing controller for the agent operating within a switched system framework. The objective is to facilitate a trajectory design that alternates between the regions $\mathcal{X}_a$ and $\mathcal{X}_u$ while collecting measurements from $\mathcal{X}_u$.

To facilitate the observer design, it is assumed that the agent is initialized inside $\mathcal{X}_a$ at time $t_0^a$. Let $t_0^u$ denote the first time instance the agent crosses the GPS boundary and enters $\mathcal{X}_u$. Subsequent boundary crossing times are indexed by $i$, such as $t_i^a$ and $t_i^u$, where $i \in \mathbb{N}$. During the time interval $t \in [t_i^a, t_i^u)$, the agent operates in $\mathcal{X}_a$, while for $t \in [t_i^u, t_{i+1}^a)$, the agent operates in $\mathcal{X}_u$. The time intervals $\Delta t_i^a := t_i^u - t_i^a$ and $\Delta t_i^u := t_{i+1}^a - t_i^u$ represent the time spent in $\mathcal{X}_a$ and $\mathcal{X}_u$ respectively.

Following the observer design for switched systems presented in [15], the state estimates are computed by solving

$$\dot{\hat{x}} = \begin{cases} f(t,\hat{x}) + u(t) + v_r, & \forall \hat{x} \in \mathcal{X}_a, \\ f(t,\hat{x}) + u(t), & \forall \hat{x} \in \mathcal{X}_u, \end{cases} \quad (12)$$

where the high-frequency sliding mode term,

$$v_r = k_1 e_1 + \bar{g}\,\mathrm{sgn}(e_1), \quad (13)$$

is introduced to enhance the observer's robustness against sudden changes in error signals. $k_1 \in \mathbb{R}^{n\times n}$ is a positive definite control gain matrix. Based on the observer dynamics in (12), a stabilizing controller can be designed as

$$u = \begin{cases} \dot{x}_d - f(t,\hat{x}) - k_2 e_2 - v_r, & \forall \hat{x} \in \mathcal{X}_a, \\ \dot{x}_d - f(t,\hat{x}(t)) - k_2 e_2, & \forall \hat{x} \in \mathcal{X}_u, \end{cases} \quad (14)$$

where $k_2 \in \mathbb{R}^{n\times n}$ is a positive definite tracking error gain matrix. The time derivative of (2) and substitution of (1) and (12) yields $\dot{e}_1$ as

$$\dot{e}_1 = \begin{cases} f(t,x) - f(t,\hat{x}) - v_r + g(t), & \forall \hat{x} \in \mathcal{X}_a, \\ f(t,x) - f(t,\hat{x}) + g(t), & \forall \hat{x} \in \mathcal{X}_u, \end{cases} \quad (15)$$

Similarly for $\dot{e}_2$, the derivative of (3), and substitution of (12) yields

$$\dot{e}_2 = \begin{cases} f(t,\hat{x}) + u + v_r - \dot{x}_d, & \forall \hat{x} \in \mathcal{X}_a, \\ f(t,\hat{x}) + u - \dot{x}_d, & \forall \hat{x} \in \mathcal{X}_u. \end{cases} \quad (16)$$

Using the controller designed in (14), the error dynamics for $e_2$ can be simplified to

$$\dot{e}_2 = -k_2 e_2, \quad \forall t \in [0, \infty). \quad (17)$$

With the error dynamics and observer update laws established, the stability of the switched system and the controller's performance in (14) can now be analyzed in the next section.

## VI. Stability Analysis for Switched Systems

Utilizing the Lyapunov function defined in (4) along with the user-defined bounds $V_u$ and $V_l$, and incorporating the error dynamics from (15) and (17), this section establishes the dwell time conditions necessary to ensure $V_l \leq V(e(t)) \leq V_u$.

*Theorem 1:* Given the error system in (15) and (17) and the candidate Lyapunov function in (4), the bounds $V_l \leq V(e(t)) \leq V_u$ holds provided the switching signal satisfies the dwell time conditions:

$$\Delta t_a^i \geq -\frac{1}{\lambda_a} \ln\left(\min\left(\frac{V_l}{V(e(t_i^a))}, 1\right)\right), \quad (18)$$

$$\Delta t_u^i \leq \frac{1}{\lambda_u} \ln\left(\frac{V_u + \frac{\bar{g}^2}{2\lambda_u}}{V(e(t_i^u)) + \frac{\bar{g}^2}{2\lambda_u}}\right), \quad (19)$$

where $\lambda_a$ and $\lambda_u$ are user-defined positive constants.

*Proof:* The orbital derivative of the candidate Lyapunov function in (4) along the error dynamics in (15) and (17) is given by

$$\dot{V}(t,e) = \begin{cases} e_1^T\left(f(t,x(t)) - f(t,\hat{x}(t)) - v_r(t) + g(t)\right) \\ + e_2^T(-k_2 e_2), & \forall \hat{x} \in \mathcal{X}_a,, \\ e_1^T\left(f(t,x(t)) - f(t,\hat{x}(t)) + g(t)\right) \\ + e_2^T(-k_2 e_2), & \forall \hat{x} \in \mathcal{X}_u,. \end{cases} \quad (20)$$

*1) Analysis in $\mathcal{X}_a$ (GPS-Available Region):* For $t \in [t_i^a, t_i^u)$, we can measure $e_1$ and hence we can compute $v_r$. Substituting $v_r$ in above equation, we get

$$\dot{V}(t,e) = e_1^T\left(f(t,x) - f(t,\hat{x}) - k_1 e_1 - \bar{g}\,\mathrm{sgn}(e_1) + g(t)\right) - e_2^T k_2 e_2.$$

Since $f$ is Lipschitz continuous in $t$, we can write

$$\|f(t,x) - f(t,\hat{x})\| \leq L_f \|e_1\|, \quad \forall x, \hat{x} \in \mathcal{X}$$

where $L_f$ is the Lipschitz constant. Substituting this, we get

$$\dot{V}(t,e) \leq L_f \|e_1\|^2 - e_1^T k_1 e_1 - e_1^T \bar{g} + e_1^T g(t) - e_2^T k_2 e_2.$$

Since $\|g(t)\| \leq \bar{g}$, we have $e_1^T g(t) \leq \bar{g}\|e_1\|, \forall t$. The terms $-\bar{g}\|e_1\| + \bar{g}\|e_1\|$ cancel out, resulting in the bound

$$\dot{V}(t,e) \leq L_f \|e_1\|^2 - e_1^T k_1 e_1 - e_2^T k_2 e_2.$$

Let $\lambda_{\min}(k_1)$ and $\lambda_{\min}(k_2)$ denote the minimum eigenvalues of $k_1$ and $k_2$, respectively. Then,

$$\dot{V}(t,e) \leq (L_f - \lambda_{\min}(k_1))\|e_1\|^2 \\ - \lambda_{\min}(k_2)\|e_2\|^2. \quad (21)$$

The above equation can be simplified to

$$\dot{V}(t,e) \leq -\lambda_a V(e),$$

where $\lambda_a \triangleq 2\min(\lambda_{\min}(k_1) - L_f, \lambda_{\min}(k_2)) \in \mathbb{R}_{>0}$. The solution of the above differential equation over the GPS available time intervals is given as

$$V(e(t)) \leq V(e(t_i^a))e^{-\lambda_a(t-t_i^a)}, \quad \hat{x} \in \mathcal{X}_a. \quad (22)$$

*2) Analysis in $\mathcal{X}_u$ (GPS-Denied Region):* For $t \in [t_i^u, t_{i+1}^a)$, the term $v_r$ is inactive. Thus,

$$\dot{V}(t,e) = e_1^T \left(f(t,x(t)) - f(t,\hat{x}(t)) + g(t)\right) - e_2^T k_2 e_2.$$

Using the Lipschitz continuity of $f$ and the bound on $g$, we have

$$\dot{V}(t,e) \leq L_f \|e_1\|^2 + \bar{g}\|e_1\| - e_2^T k_2 e_2.$$

Above inequality can be simplified as

$$\dot{V}(t,e) \leq L_f \|e_1\|^2 + \bar{g}\|e_1\| - \lambda_{\min}(k_2)\|e_2\|^2. \quad (23)$$

Choosing $\lambda_u \triangleq 2L_f + 1$, we can write

$$\dot{V}(t,e) \leq \lambda_u V(e(t)) + \frac{\bar{g}^2}{2},$$

The solution of the above differential equation on the interval $[t, t_i^u)$ yields

$$V(e(t)) \leq V(e(t_i^u))e^{\lambda_u(t-t_i^u)} \\ + \frac{\bar{g}^2}{2\lambda_u}\left(e^{\lambda_u(t-t_i^u)} - 1\right), \forall \hat{x} \in \mathcal{X}_u. \quad (24)$$

To derive the desired dwell time $\Delta t_i^a$ in $\mathcal{X}_a$, we utilize the solution in (22) at the end of the interval $[t_i^a, t_i^u)$ such that $t = t_i^u$ yields

$$V_l \leq V(e(t_i^a))e^{-\lambda_a \Delta t_i^a}.$$

Rearranging for $\Delta t_i^a$, we want

$$\Delta t_i^a \geq -\frac{1}{\lambda_a} \ln\left(\frac{V_l}{V(e(t_i^a))}\right). \quad (25)$$

Similarly, to get the maximum dwell time $\Delta t_i^u$ in $\mathcal{X}_u$, we utilize the solution (24) at the end of the interval $[t_i^u, t_{i+1}^a)$, such that $t = t_{i+1}^a$ yields

$$V(e(t_i^u))e^{\lambda_u \Delta t_i^u} + \frac{\bar{g}^2}{2\lambda_u}\left(e^{\lambda_u \Delta t_i^u} - 1\right) \leq V_u.$$

Rearranging for $\Delta t_i^u$ yields

$$\Delta t_i^u \leq \frac{1}{\lambda_u} \ln\left(\frac{V_u + \frac{\bar{g}^2}{2\lambda_u}}{V(e(t_i^u)) + \frac{\bar{g}^2}{2\lambda_u}}\right). \quad (26)$$

which completes the proof for desired dwell-time conditions. ∎

In the region $\mathcal{X}_a$, the exponential decay of the Lyapunov function indicates a stable behavior governed by the designed error system dynamics. Conversely, in the region $\mathcal{X}_u$, the Lyapunov function exhibits growth due to the loss of state feedback. The dwell-time conditions play a crucial role in designing different segments of $\bar{x}_d$ in $\mathcal{X}_a$ and $\mathcal{X}_u$ while ensuring boundedness of error trajectories.

## VII. SWITCHING TRAJECTORY DESIGN

Using the dwell time conditions derived in the last section, the switching trajectory $\bar{x}_d(t)$ can now be developed. To facilitate the design, we assume a path $x_d$ that connects all the measurement locations in $\mathcal{X}_u$. Therefore, our designed switching trajectory $\bar{x}_d$ should intermittently overlap $x_d$ to collect field measurement along $x_d$ while switching between $\mathcal{X}_u$ and $\mathcal{X}_a$ regions. An orthogonal projection vector $x_b \in \mathbb{R}^n$ can now be defined such that it connects the points on the path $x_d$ to the closest point on the boundary of $\mathcal{X}_a$, i.e.,

$$x_b = \arg\min_{y \in \partial \mathcal{X}_a} \{\|y - x_d\|\} \quad (27)$$

Given projection vectors, our dwell time $\Delta t_i^u$ in $\mathcal{X}_u$ can now be divided into three segments. In the first time segment ($t_i^u \leq t < t_i^{u1}$) the agent covers $x_b$ distance to join $x_d$, in the second time segment ($t_i^{u1} \leq t < t_i^{u2}$), $\bar{x}_d$ follows $x_d$ to collect measurements and in the third time segment ($t_i^{u3} \leq t < t_i^{u3}$) the agent travels back to $\mathcal{X}_a$ to attain state feedback. The parameters $\rho_i^a, \rho_i^{u1}, \rho_i^{u2}, \rho_i^{u3}$ are used partition the dwell times as

$$\rho_i^a(t) = \frac{t - t_i^a}{\Delta t_i^a}, \quad \rho_i^{uj+1}(t) = \frac{t - \left(t_i^u + \sum_{k=0}^{j} w_k \Delta t_i^u\right)}{w_{j+1} \Delta t_i^u}, \quad (28)$$

where $j \in \{0, 1, 2\}$ is the index for partitioning and $w_j \in [0, 1)$ are the weights used to partition the dwell time such that $\sum_{j=0}^{2} w_j = 1$. The final partition $t_i^{u3}$ coincides with $t_i^{a+1}$ to connect the next cycle of the switching trajectory.

To facilitate smooth transitions between different segments of the switching trajectory, the smoother step function, as defined in [17], is utilized.

$$S(\rho) = 6\rho^5 - 15\rho^4 + 10\rho^3, \quad (29)$$

where the user defined parameter $\rho \in [0, 1]$ controls the smoothness of the function. The switching trajectory can now be introduced as

$$\bar{x}_d(t) := \begin{cases} H\left(S(\rho_i^{u1}), g(x_d, t), x_b(t)\right), & t_i^u \leq t < t_i^{u1} \\ H\left(S(\rho_i^{u2}), g(x_d, t), g(x_d, t)\right), & t_i^{u1} \leq t < t_i^{u2} \\ H\left(S(\rho_i^{u3}), x_\epsilon(t), g(x_d, t)\right), & t_i^{u2} \leq t < t_i^{u3}, \\ H\left(S(\rho_i^a), x_b(t), x_\epsilon(t)\right), & t_{i+1}^a \leq t < t_{i+1}^u \end{cases} \quad (30)$$

where the interpolation function $H(S(\cdot), q(t), r(t))$ is defined as

$$H(S(\cdot), q(t), r(t)) := S(\cdot)q(t) + [1 - S(\cdot)]r(t), \quad (31)$$

for $q(t), r(t) \in \mathbb{R}^n$. The function $g : \mathbb{R}^n \times \mathbb{R} \to \mathbb{R}^n$ computes the state that the agent should ideally be in at time $t$ along the desired path $x_d$. In the third time segment $x_\epsilon(t)$ indicate the cushion trajectory $x_\epsilon(t) \in \mathbb{R}^n$ as

$$x_\epsilon(t) \triangleq x_b(t) + \phi(t), \quad (32)$$

where $\phi(t) \in \mathbb{R}^n$ and $\|\phi(t)\| \geq 2\sqrt{V_u}$.

The dwell time analysis in the previous section indicates that the agent must stay for at least $\Delta t_i^a$ seconds in $\mathcal{X}_a$ and at most $\Delta t_i^u$ seconds in $\mathcal{X}_u$ to satisfy the error bounds. One could design a trajectory that spends exactly $\Delta t_i^u$ in $\mathcal{X}_u$; however, due to unmodeled dynamics, the agent may end up spending more time in $\mathcal{X}_u$ than allowed. To mitigate this error due to unmodeled uncertainty, we elongate the trajectory $x_b$ with $\phi(t)$ in the last time segment ($t_i^{u2} \leq t < t_i^{u3}$) so that the agent is guaranteed to attain state feedback even with maximum error ($V_u$).

The switching trajectory designed in this section ensures that the agent visits all measurement locations in GPS-denied regions without violating the desired error bounds. The measurements are collected while following $\bar{x}_d$ and are used to train a GP model using (10) and (11).

## VIII. SIMULATION STUDY

A simulation study is performed to evaluate the performance of the proposed methodology. The scalar field model is generated by multiple field sources using a Gaussian decay function, given as

$$h(\mathbf{x}) = \sum_{i=1}^{N} I_i \exp\left(-\gamma \left((x - x_i)^2 + (y - y_i)^2\right)\right), \quad (33)$$

where $(x, y) \in [-5, 5]$ represents the domain of the scalar field, $(x_i, y_i)$ are the coordinates of the sources, $I_i$ represents the intensity of each source, and $\gamma$ is a decay constant that controls the spread of each source's influence.

The intensity values for the field sources are $I_1 = 5$, $I_2 = 5$, $I_3 = 4$, and $I_4 = 4$, with locations $(-2, 0)$, $(2, 0)$, $(0, 2)$, and $(0, -2)$ respectively. The decay constant is set to $\gamma = 0.5$, ensuring a smooth decrease in field intensity with increasing distance from each source. The resulting scalar field is shown in Figure 1 with a 3D surface plot of the scalar field.

The agent dynamics is assumed to be $f(t, x) = Ax$, where $A = 0.5I_3$, and the unmodeled dynamics $g(t)$ are chosen from a uniform distribution $[0, 0.06]$. The initial states for the agent, representing the position coordinates and the orientation angle, are selected as $x(0) = [0.1, 0.2, 0]$. The state estimates are selected as $\hat{x}(0) = [0.2, 0.3, \frac{\pi}{6}]$. The GPS boundary is assumed to be at a radius of 1m, and the measurement locations required to map the field are assumed to lie on the desired trajectories $x_d$ in $\mathcal{X}_u$ at a radius of 2 meters centered at same origin as GPS boundary as shown in Figure (2). The observer and the controller gains are chosen as $k_1 = 3I_3$ and $k_2 = 3I_3$, respectively. The Lyapunov function bounds are set to $V_u = 0.2025$ and $V_l = 1 \times 10^{-4}$.

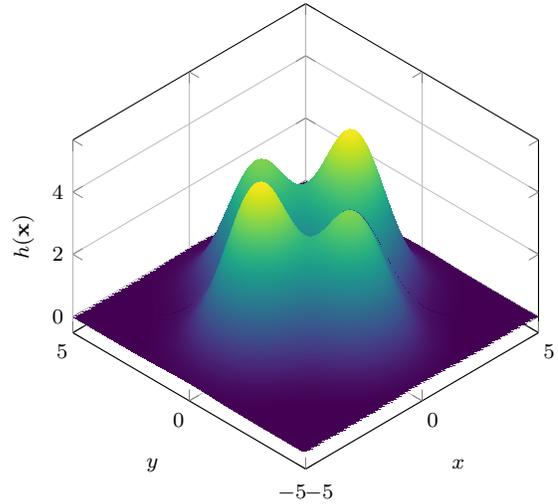

Fig. 1: 3D Surface Plot of the true scalar field $g$ used in the simulation.

The partition weights used to design the switching trajectory $\bar{x}_d(t)$ according to the design in Section VII, are chosen as $w_0 = 0.2$, $w_1 = 0.6$, and $w_2 = 0.2$. The switching trajectory follows the design described, based on the dwell time constraints, switching between $\mathcal{X}_a$ and $\mathcal{X}_u$ to complete the full 360 degrees of $x_{d1}$ and collect all field measurement required to map the field.

The trajectory of actual states $x(t)$ and estimated states $\hat{x}(t)$ are shown in Figure 2. The robot starts near the origin and follows the x-axis to join $x_d$. The desired circular trajectory is also shown in the plot. While following $x_d$, the agent also collects measurements for the GP model, as shown in Figure 5. Plot of the trajectory tracking error $e_w(t)$ is shown in Figure 4. The scalar field predicted by the two GP models is presented in Figure 3. The root mean squared value of the error between actual field values and GP predicted values at all testing locations, is shown in Figure 6.

## IX. DISCUSSION

This study integrates the switched systems framework for mapping scalar fields in GPS-denied environments. However, no state feedback constrains the agent's range of travel in $\mathcal{X}_u$. To mitigate this, a switched system approach can be employed to enable the agent to visit measurement locations in GPS-denied areas and provide an accurate field map.

Future work aims to incorporate disturbance observers to facilitate its application to more complex topological structures. Furthermore, the impact of localization errors on the Gaussian Process (GP) model needs to be investigated, as the GP model is trained on estimated states rather than actual states. Position estimation uncertainties may propagate through the GP training process, leading to deviations from the true scalar field representation. To address this challenge, a noisy-input Gaussian Process framework will be utilized to construct a more accurate and robust field map.

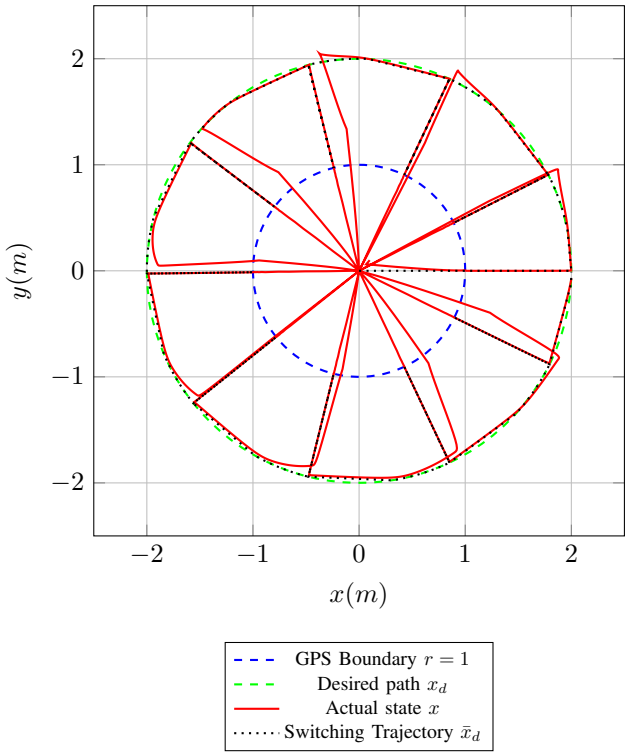

Fig. 2: A plot of the motion of the agent over the part of the experiment where the agent is confined to measurement locations on $x_d$, superimposed on the path $x_d$ and the desired trajectory $\bar{x}_d$

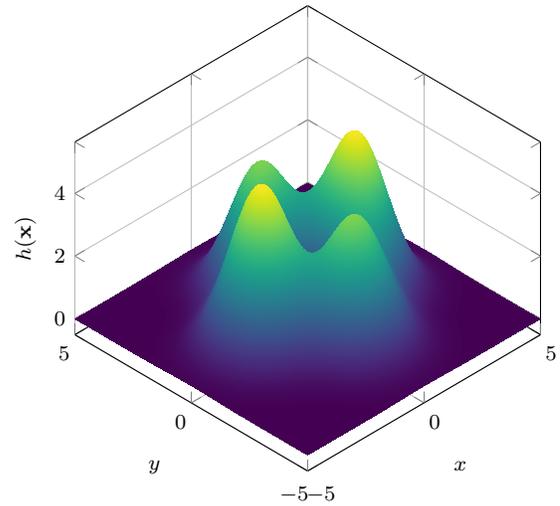

Fig. 3: The mean function of the GP predicted scalar field is plotted in this figure.

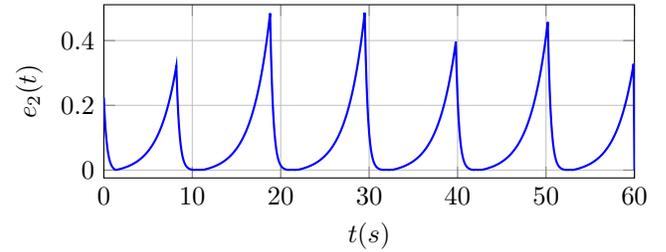

Fig. 4: The state estimation error is plotted as a function of time.

## X. Conclusion

This paper presents a switched systems framework for autonomous navigation and scalar field mapping in GPS-denied environments. By leveraging intermittent state measurements from GPS-enabled regions, the proposed method ensures accurate state estimation while mitigating error accumulation due to the absence of continuous feedback.

Simulation results demonstrate the approach's effectiveness in achieving reliable field mapping despite localization uncertainties. The integration of Gaussian Process (GP) regression enhances sparse measurement interpolation and provides uncertainty quantification, improving the robustness and efficiency of the mapping process.

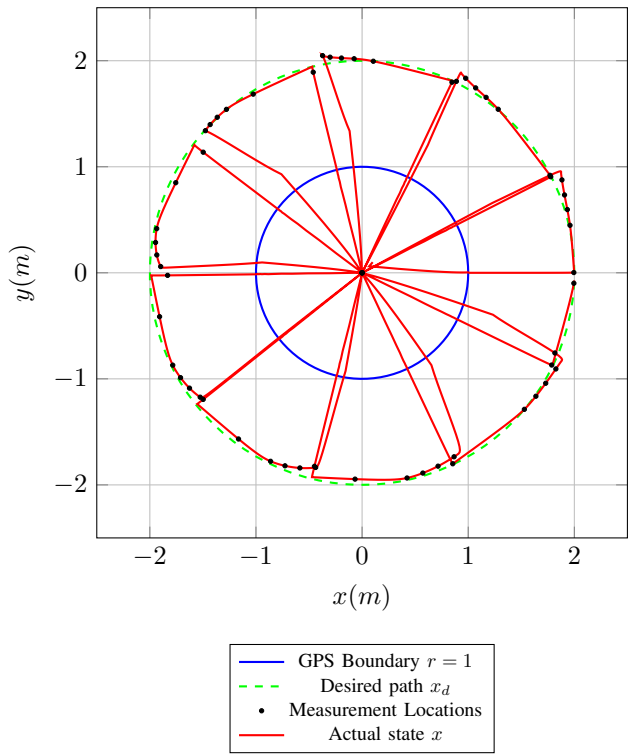

Fig. 5: The measurement locations used to train the GP models are shown as black dots on the top view of the scalar field.

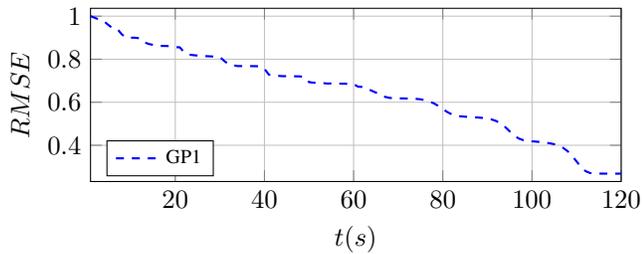

Fig. 6: A Comparison of the normalized RMSE generated by the two GP models.